\documentclass[review]{elsarticle}
\usepackage{amsmath}
\usepackage{amsfonts}
\usepackage{lineno,hyperref}
\usepackage{xcolor}
\usepackage{natbib}
\modulolinenumbers[5]

\journal{Journal of \LaTeX\ Templates}

\begin{document}

\begin{frontmatter}

\title{ Fast Stable Parameter Estimation for Linear Dynamical Systems}

\author[mymainaddress]{Carey, M.\corref{mycorrespondingauthor}}
\ead[url]{https://data2dynamics.ucd.ie/}
\author[Jimsaddress]{Ramsay, J. O. }

\cortext[mycorrespondingauthor]{michelle.carey@ucd.ie}
\address[mymainaddress]{School of Mathematics and Statistics, University College Dublin, Dublin, 
Ireland.}
\address[Jimsaddress]{Department of Mathematics and Statistics, McGill University, Montr\'eal, 
Qu\'ebec, Canada.}

\begin{abstract}
Dynamical systems describe the changes in processes that arise naturally from their underlying physical 
principles, such as the laws of motion or the conservation of mass, energy or momentum. These models 
facilitate a causal explanation for the drivers and impediments of the processes. But do they describe 
the behaviour of the observed data? And how can we quantify the models' parameters that cannot be 
measured directly? This paper addresses these two questions by providing a methodology for 
estimating the solution; and the parameters of linear dynamical systems from incomplete and noisy 
observations of the processes. 

The proposed procedure builds on the parameter cascading approach, where a linear combination of 
basis functions approximates the implicitly defined solution of the dynamical system. The systems' 
parameters are then estimated so that this approximating solution adheres to the data. By taking 
advantage of the linearity of the system, we have simplified the parameter cascading estimation 
procedure, and by developing a new iterative scheme, we achieve fast and stable computation. 

We illustrate our approach by obtaining a linear differential equation that represents real data from 
biomechanics. Comparing our approach with popular methods for estimating the parameters of linear 
dynamical systems, namely, the non-linear least-squares approach, simulated annealing, parameter 
cascading and smooth functional tempering reveals a considerable reduction in computation and an 
improved bias and sampling variance. 
\end{abstract}


\end{frontmatter}

%\linenumbers

\section{Introduction}

\begin{figure}[!h]
	\centering
	\caption{The circles illustrate the accelerometer readings of the head acceleration before and after a 
		blow to the cranium of a cadaver. The dashed line represents a 
		unit pulse function which denotes the blow to the skull. This function initiates at 14 milliseconds 
		and lasts for 1 millisecond.\label{Head_Imp1}}
	\includegraphics[width=.7\textwidth]{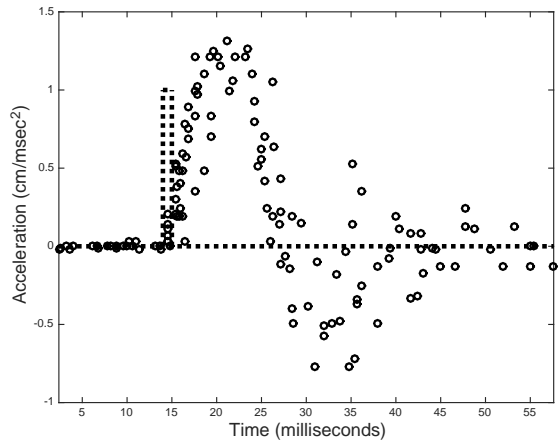}
\end{figure}

Dynamical systems typically translate the natural phenomena into a set of equations based on the 
motion or equilibrium of the system as determined by its mechanics, chemistry, biology, etc. These 
models explain the underlying mechanisms that drive or hinder a processes behaviour. A set of linear 
differential equations denotes a linear dynamical system.  Let the $p^{th}$ derivative of the function $x$ 
at time $t$ be $D^{p}x(t).$ A $p^{th}$ order differential equation specifies how the behaviour of the 
$p^{th}$ derivative depends on the lower order derivatives, $D^{0}x(t),\ldots,D^{p-1}x(t),$ and other 
external variables, $u_{1}(t),\ldots,u_{Q}(t),$ that is,
\begin{equation}\label{ODE}
D^{p}x(t) = - \sum_{r=0}^{p-1} \beta_{r}(t|\boldsymbol{\theta}) D^{r}x(t) + \sum_{q=1}^{Q} 
\alpha_{q}(t|\boldsymbol{\theta}) u_{q}(t), 
\end{equation}
where $t \in [t_{1}, t_{N}],$ the coefficient functions $\beta_{r}(t|\boldsymbol{\theta})$ and 
$\alpha_{q}(t|\boldsymbol{\theta})$ are functions of $t$ that are dependent on a vector of parameters 
$\boldsymbol{\theta}$ and $u_{q}(t)$ is the $q^{th}$ function at time $t$ representing the $q^{th}$ 
external variable. The differential equation is linear if the functions $\beta_{r}(t|\boldsymbol{\theta})$, 
$\alpha_{q}(t|\boldsymbol{\theta})$ and $u_{q}(t)$ do not depend on the values of $x.$\footnote{For 
simplicity of notation hereafter we will work with the single $p^{th}$ order linear differential equation in 
(\ref{ODE}). Although the extension to a set of $p^{th}$ order linear ODEs in a dynamical system is 
trivial.} This formulation encompasses a broad range of phenomena, including those observed in climate 
science, biology and ecology. See for example, 
\cite{tung2007topics,jones2009differential,jopp2010modelling} and the references therein. The main challenge is determining the values of the parameters $\boldsymbol{\theta}$, defining 
$\beta_{r}(t|\boldsymbol{\theta})$ and  $\alpha_{q}(t|\boldsymbol{\theta})$  in (\ref{ODE}), that ensure 
the approximating solution of (\ref{ODE}) evaluated at the observed times, adheres to the observed behaviour of the process.  We illustrate this problem by presenting an example of a linear differential equation for modelling head acceleration. Figure (\ref{Head_Imp1}) depicts $133$ observations of head acceleration (in cm/msec$^2$) measured 14 milliseconds before and 42.6 milliseconds after a blow to 
the cranium. The dashed line represents the unit pulse function which denotes the strike to the cranium 
that lasted one millisecond. The experiment, a simulated motor-cycle crash, is described in detail in 
\cite{schmidt1981biomechanical}. 
%%%%%%%%%%%%%%%%%%%%%%%%%%%%%%%%%%%%%%%%%%%%%%%%%%%%%%%%%%%%%%%%%%%%%%%%%%%%%%%%%%%%%%%%%%%%%%%%%%%%%%%%%%%%%%%%%%%%%%%%%%%%%%%%%%%%%%%%%%%%%%%%%%%%%%%%%%%%%%%%%%%%%%%%%%%%%%%%%%%%%%%%%%%%%
% 
%%%%%%%%%%%%%%%%%%%%%%%%%%%%%%%%%%%%%%%%%%%%%%%%%%%%%%%%%%%%%%%%%%%%%%%%%%%%%%%%%%%%%%%%%%%%%%%%%%%%%%%%%%%%%%%%%%%%%%%%%%%%%%%%%%%%%%%%%%%%%%%%%%%%%%%%%%%%%%%%%%%%%%%%%%%%%%%%%%%%%%%%%%%%%
Mechanical principles imply that the acceleration $x(t)$ can be modelled by a second order linear 
differential equation with a unit pulse external function $u(t)$ representing the blow to the cranium, as 
shown by the dashed lines in Figure (\ref{Head_Imp1}). The three parameters $\beta_{0}$, $\beta_{1} $ 
and $\alpha$ in 
\begin{equation}\label{HIE}
D^2x(t) =  -\beta_{0} x(t) - \beta_{1}  Dx(t) + \alpha  u(t),
\end{equation}
convey the period of the oscillation, the change in its amplitude, as $t \rightarrow \infty$ the oscillations 
decay exponentially to zero, and the size of the impact from the unit pulse respectively. Our objective is 
to estimate the acceleration $x$ and the parameters $\boldsymbol{\theta}=[\beta_{0},\beta_{1},\alpha]$ 
in (\ref{HIE}) so that the approximated solution of (\ref{HIE}), $\hat{\textbf{x}}(\textbf{t}|\boldsymbol{\boldsymbol{\theta}})$, evaluated at the observed times, 
$\textbf{t}=[t_{1},\ldots,t_{N}]$, adheres 
to the data in Figure (\ref{Head_Imp1}).

Non-linear least squares (NLS) is the most common approach 
\citep{hemker1972numerical,bard1974nonlinear,schittkowski2002dynamics} for estimating the solution 
$x(t)$ and the parameters $\boldsymbol{\theta}$ for the differential equation in (\ref{ODE}). Given a set 
of initial conditions, $D^r x(0)$ for $r=0,\ldots,p-1$, and a period of the domain over which the solution 
is sought, $[t_{1},t_{N}]$, the solution of (\ref{ODE}) can be approximated by a numerical iterative 
method (e.g. Runge-Kutta methods). NLS then estimates the parameters $\boldsymbol{\theta}$ by 
minimising the difference between the approximated numerical solution of (\ref{ODE}) and the observed 
data values. Typically this minimisation problem has many local minima. 
As shown in \cite{gonzalez2006parameter} simulated annealing (SA), introduced by 
\cite{kirkpatrick1983optimization}, can be used to overcome the topological difficulties in the minimisation problem. The NLS and SA approaches are both computationally intensive as a numerical 
approximation to the solution of the differential equation in (\ref{ODE}) is required for each update of 
$\boldsymbol{\theta}$. Additionally, the initial values $D^r x(0)$ for $r=0,\ldots,p-1$ are not usually 
available in exact form. Therefore, we often need to minimise the objective function with respect to 
$\boldsymbol{\theta}$ and $D^r x(0)$ for $r=0,\ldots,p-1$. This adds a great deal of extra computation 
and complexity to the optimisation. Parameter cascading (PC) attributable to 
\cite{ramsay2007parameter} alleviates the computational cost associated with repeatedly numerically 
solving the differential equation and does not require the initial values $D^r x(0)$ for $r=0,\ldots,p-1$ to 
be available in exact form. PC uses a linear combination of basis functions to approximate the 
solution of (\ref{ODE}) and the estimated parameters $\hat{\boldsymbol{\theta}}$ are obtained by 
ensuring that the approximating basis function expansion adheres to the data. Similar to NLS, PC has 
topological difficulties in minimising the data misfit. Smooth functional tempering (SFT) proposed by 
\cite{Campbell2012} implements a Bayesian version of PC and borrows insights from parallel tempering \cite{geyer1991markov,falcioni1999biased}
to overcome the topological difficulties. SFT produces accurate estimates of 
$\boldsymbol{\theta}$, but it is very computationally expensive. Quick and easy procedures have 
been proposed by \cite{varah1982spline,liang2008parameter,brunel2008parameter,Hall} and 
\cite{GenPen}. These methods do not account for the hierarchical structure of the parameters. The 
parameters that approximate the solution of (\ref{ODE}) are dependent on the parameters $\boldsymbol{\theta}$, which determine the shape of the solution. As a consequence, each 
method reports a considerable increase in the bias of $\hat{\boldsymbol{\theta}}$ when compared to 
the estimates produced by the  SFT, PC, SA or NLS approaches.

Motivated by the drawbacks of the existing methods discussed above, we introduce a version of the PC 
approach called data to linear dynamics ``Data2LD". Data2LD is a fast and stable version of the PC 
approach for estimating the parameters of linear dynamical systems. First, we reduce the complexity of 
the PC estimation procedure, which has the advantages of speed and ease of use. Then analogous to 
SA and SFT, we propose an iterative scheme to overcome the topological difficulties in minimising the 
data misfit. One of the primary benefits of this algorithm is that it facilitates an accurate and stable 
estimation of the solution, $x(t),$ and the parameters, $\boldsymbol{\theta}$ defining 
$\beta_{r}(t|\boldsymbol{\theta})$ and $\alpha_{q}(t|\boldsymbol{\theta})$ in (\ref{ODE}). In comparison 
to other techniques, namely NLS, SA, PC and SFT our proposed method benefits from estimates of 
$\boldsymbol{\theta}$ and $x(t),$ with an improved bias and sampling variance obtained at a fraction of 
the computational cost.

Section (\ref{Background}) briefly reviews the existing approaches for estimating the solution $x(t)$ 
and the 
parameters $\boldsymbol{\theta}$ from data. Section (\ref{Data2LD}) describes our approach detailing 
the dynamic model-fitting criteria, proposing an iterative scheme for estimating $\boldsymbol{\theta}$ 
and providing formulae for approximating the sampling variance of $\boldsymbol{\theta}$ and $x(t)$. 
Section (\ref{subsec:TBI}) illustrates the estimation of the solution and the parameters 
$\boldsymbol{\theta}$ from noisy incomplete data by obtaining a linear differential equation for 
modelling head acceleration. Section (\ref{sec:simulation}) presents a simulated data example and its 
performance.

\section{Background}\label{Background}

For a detailed account of modern methods for estimating parameters in linear and non-linear differential equations see \cite{ramsay2017dynamic}. Sections (\ref{NLSD}) to (\ref{S_SFT}) briefly outlines four popular approaches for estimating the 
solution $x(t)$ and the parameters $\boldsymbol{\theta}$ of the differential equation in (\ref{ODE}). 

\subsection{Non-linear least squares (NLS)}\label{NLSD}

Given an initial estimate $\boldsymbol{\theta}_{0}$ of $\boldsymbol{\theta}$, and a set of $p$ initial 
values, $D^r x(0),\, r=0,\ldots,p-1$, a numerical approximation to the solution of the differential 
equation in (\ref{ODE}), $\hat{x}(\textbf{t},\boldsymbol{\theta}_{0},D^{0}x(0), \ldots,D^{p-1}x(0)),$ 
evaluated at the observed times $\textbf{t}$, is computed using a method for initial value problems such 
as a Runge-Kutta method. The estimated parameters are then obtained by minimising, 
\begin{equation}\label{NLS}
\hat{\boldsymbol{\theta}} = 
\underset{\boldsymbol{\theta}}{\textrm{min}} \sum_{i=1}^{N} 
\left[y_{i}-\hat{x}(t_{i},\boldsymbol{\theta},D^{0}x(0), \ldots, D^{p-1}x(0) )\right]^{2},
\end{equation}
with the initial condition $\boldsymbol{\theta}=\boldsymbol{\theta}_{0},$ using a gradient-based 
optimisation method (e.g.  the trust-region-reflective algorithm). Typically, one obtains the gradient and 
hessian of (\ref{NLS}) evaluated at the current estimate of $\boldsymbol{\theta}$ using numerical 
differentiation (e.g. finite difference approximations). Often, the topology of the objective function in 
(\ref{NLS}) is undesirable with local minima, ridges, ripples and large flat segments. 

\subsection{Simulated annealing (SA)}\label{Sim_A}

The Metropolis-Hastings (MH) algorithm draws samples of 
$\boldsymbol{\theta}$ 
from the conditional distribution 
\begin{equation}\label{SANLS}
\pi(\boldsymbol{\theta} | T, \textbf{y}, \textbf{t}, D^{0}x(0), \ldots, D^{p-1}x(0)) = \exp 
\left(\frac{-\sum_{i=1}^{N} 
	\left[y_{i}-\hat{x}(t_{i},\boldsymbol{\theta},D^{0}x(0), \ldots, D^{p-1}x(0) )\right]^{2}}{T}\right),
\end{equation}
for a fixed temperature $T$. As $T\rightarrow \infty$, $\pi$ tends to a uniform probability density 
function and as $T\rightarrow 0$, $\pi$ tends to a delta function located at the global maximum of 
(\ref{SANLS}), which is equivalent to the global minimum of (\ref{NLS}).  A temperature ladder $T_i$ for 
$i=1,\ldots,M$, is constructed and the MH algorithm progressively samples from the conditional 
distribution in (\ref{SANLS}) as the temperature is adjusted from high to low values. At the $u^{th}$ 
iteration if the value of $\pi^{u}$ is greater than $\pi^{u-1},$ the new $\hat{\boldsymbol{\theta}}^{u}$ is 
accepted. Otherwise, the new $\hat{\boldsymbol{\theta}}^{u}$ is accepted at random with a probability 
$1/(1+\exp((\pi^{u-1}-\pi^{u})/T_{i}).$ A smaller temperature or a larger distance between $\pi^{u-1}$ and 
$\pi^{u}$ will lead to a smaller acceptance probability. SA provides a means to escape local optima by 
accepting steps which decrease $\pi$ in hopes of finding a global optimum. 

%The performance of the algorithm is critically dependent on the choice of the temperature 
%ladder see \cite{stander1994temperature} for details. 

\subsection{Parameter cascading (PC)}\label{PC_S}

PC approximates the solution of (\ref{ODE}) by a linear combination of basis functions, $
\textbf{x} \approx \sum_{k=1}^{K}c_{k}\phi_{k}(\textbf{t})= 
\boldsymbol{\Phi}\textbf{c},
$where $\boldsymbol{\Phi}$ is the $N \times K$  matrix containing the basis function $\phi_{k}(t)$ 
evaluated at the locations $\textbf{t}$ and $\boldsymbol{c}$ is a vector of length $K$ containing the 
corresponding coefficients. PC defines the coefficient $c_{k}$ as a smooth function of the parameters 
$\boldsymbol{\theta}$ and $\lambda,$ where $\lambda$ is a regularity parameter that determines the 
trade-off between $\textbf{x}'s$ fit to the data and adherence to the differential equation in (\ref{ODE}). 
For fixed $\lambda$, the estimated parameters $\hat{\textbf{c}}(\boldsymbol{\theta},\lambda),$ are 
produced by minimising
\begin{equation}\label{c}
J(\textbf{c}| \boldsymbol{\theta},\lambda) = 
\left[ 
\textbf{y}-\boldsymbol{\Phi}\textbf{c}
\right]^{T}\left[ 
\textbf{y}-\boldsymbol{\Phi}\textbf{c} \right] + \lambda 
\textbf{L}(\boldsymbol{\theta},\boldsymbol{\Phi}\textbf{c}),
\end{equation}
with respect to $\boldsymbol{c}$ each time 
the parameter vector $\boldsymbol{\theta}$ is updated. The penalty term, 
$\textbf{L}(\boldsymbol{\theta},\boldsymbol{\Phi}\textbf{c}),$ in (\ref{c}) is the square 
$\mathbb{L}_{2}$ norm of the differential equation in (\ref{ODE}) with $x$ replaced by 
$\boldsymbol{\Phi}\textbf{c}$. The 
parameters, $\boldsymbol{\theta},$ for fixed $\lambda$ are then estimated so the resulting 
approximating solution, $\hat{\textbf{x}}$, adheres to the data, which is achieved by minimising
\begin{equation}\label{theta}
H(\boldsymbol{\theta}| \lambda) = 
\left[ 
\textbf{y}-\boldsymbol{\Phi}\hat{\textbf{c}}(\boldsymbol{\theta},\lambda) 
\right]^{T}\left[ 
\textbf{y}-\boldsymbol{\Phi}\hat{\textbf{c}}(\boldsymbol{\theta},\lambda) 
\right],
\end{equation}
with respect to $\boldsymbol{\theta}$. The regularity parameter $\lambda$ is typically chosen by minimising generalized cross validation. Similar to NLS, PC has topological difficulties in the minimisation 
of (\ref{theta}). 

\subsection{Smooth Functional Tempering (SFT)}\label{S_SFT}

SFT implements a Bayesian version of PC for fixed $\lambda$,
\begin{eqnarray}\label{SFT}
\nonumber
\pi( \textbf{y}| \textbf{x},\sigma^2   ) &=& 
\frac{1}{(2\pi)^{\frac{N}{2}}\sigma^{N}} \exp \left(-\frac{1}{2\sigma^{2}} 
 \left[ 
 \textbf{y}-\textbf{x}
 \right]^{T}\left[ 
 \textbf{y}-\textbf{x} \right] \right),\\
 \nonumber
\pi( \textbf{x}| \boldsymbol{\theta},\lambda) &=& 
 \exp \left(-\frac{\lambda}{2}\textbf{L}(\boldsymbol{\theta},\textbf{x}) 
\right),\\
\pi(\boldsymbol{\theta}|\textbf{y}) &\propto & \pi( \textbf{y}| \textbf{x},\sigma^2)\pi( \textbf{x}| 
\boldsymbol{\theta},\lambda) \pi(\boldsymbol{\theta}) \pi(\sigma^2),
\end{eqnarray}
where $\pi(\boldsymbol{\theta})$ and $\pi(\sigma^2)$ are prior distributions defined for 
$\boldsymbol{\theta}$ and $\sigma^2$ respectively. Similar to PC, the sampling problem becomes difficult due to the multi-modality of the posterior surface 
$\pi(\boldsymbol{\theta}|\textbf{y})$. Gaps between modes can be 
traversed at lower values of $\lambda$, while individual modes can be 
efficiently explored at higher values of $\lambda$. In contrast  
to SA, SFT replaces the unidirectional reduction of the temperature ladder by a set of concurrent 
simulations at $M$ different temperatures $\{\lambda_m | m=1,\ldots,M\}$. At the $u^{th}$ iteration, 
each of the $M$ chains independently performs a MH step to update 
$\boldsymbol{\theta}_{1}^{u},\ldots,\boldsymbol{\theta}_{M}^{u}$. Let $U$ be a 
uniform random variable on $[0, 1]$. If $U$ is less than a threshold value (e.g. $0.5$) then a randomly 
selected neighbouring pair of chains, refereed to as $m$ and $m+1$, exchange states, that is, 
$\boldsymbol{\theta}_{m}^{u} \rightarrow 
\boldsymbol{\theta}_{m+1}^{u}$ and $\boldsymbol{\theta}_{m+1}^{u} \rightarrow 
\boldsymbol{\theta}_{m}^{u}$. This exchange is accepted with probability  
$ \textrm{min}\left(1,\frac{\pi_{m}(\boldsymbol{\theta}_{m+1}^{u}|
	\textbf{y})\pi_{m+1}(\boldsymbol{\theta}_{m}^{u}|\textbf{y})}{\pi_{m}(\boldsymbol{\theta}_{m}^{u}|\textbf{y})
\pi_{m+1}(\boldsymbol{\theta}_{m+1}^{u}|\textbf{y})}\right).$
SFT increases the efficiency of the sampling of $\pi(\boldsymbol{\theta}|\textbf{y})$ and thus can 
improve the convergence to a global minimum.

\section{Data2LD}\label{Data2LD}

Here we present a version of PC, called Data2LD, designed for the estimation of the parameters of linear 
differential equations as in (\ref{ODE}). 

\subsection{The dynamic model-fitting criterion}
\label{subsec:criteria}

Approximate the solution of the differential equation in (\ref{ODE}) by a basis function 
expansion 
\begin{equation}\label{BE2} 
x(t) \approx \sum_{k=1}^{K}c_{k}\phi_{k}(t).
\end{equation}
Assume the coefficients $c_k$ in (\ref{BE2}) are smooth functions of the parameters  
$\boldsymbol{\theta}$ and $\rho,$ that is, $c_k(\boldsymbol{\theta}, \rho),$ where $\rho$ controls 
$x$'s approximation adherence to the differential equation in (\ref{ODE}). The basis functions 
$\phi_{k}(t)$ are chosen to reflect the characteristics of the data. For example, if the data exhibit cyclical 
behaviour then Fourier basis may be desirable. We recommend B-spline basis functions due to their 
flexibility and computational efficiency. The number of basis functions must be large enough to 
guarantee that the regularisation is controlled by the choice of the regulating parameter $\rho$ and to 
ensure a satisfactory approximation of the highest order derivative $D^{p}x(t)$.  An exact 
representation or interpolation of the data is achieved when $K = N$. As advised in 
\cite{silverman2005functional}, we typically set $K=N+O-2,$ where $O$ is the order of the B-spline 
basis functions. We recommend setting the order $O>p+3$ to ensure that the highest order derivative 
$D^{p}x(t)$ is at least approximated with piece-wise cubic functions. For further details on possible 
basis functions and choices of $K$ see \cite{silverman2005functional}.

Let $\textbf{R}(\boldsymbol{\theta})$ be the $K \times 
K$ matrix with entries
\begin{eqnarray} \label{Reqtn}
\nonumber
\textbf{R}_{k,j}(\boldsymbol{\theta}) &=& \int_{t_{1}}^{t_{N}} 
\left[ D^{p}\phi_{k}(t) + 
\sum_{r=0}^{p-1} \beta_{r}(t|\boldsymbol{\theta}) D^{r} \phi_{k}(t) \right] \times \left[ D^{p}\phi_{j}(t) +
\sum_{r=0}^{p-1} \beta_{r}(t|\boldsymbol{\theta}) D^{r} \phi_{j}(t) \right] \textrm{d}t,
\end{eqnarray} 
and $\textbf{S}(\boldsymbol{\theta})$ be a $K \times 1$ vector with entries
\begin{eqnarray} \label{Seqtn}
\nonumber
\textbf{S}_{k}(\boldsymbol{\theta}) &=& \int_{t_{1}}^{t_{N}} 
\left[ D^{p}\phi_{k}(t) +
\sum_{r=0}^{p-1} \beta_r(t|\boldsymbol{\theta}) D^{r}\phi_{k}(t)\right] \times 
\left[ - \sum_{q=1}^{Q}  \alpha_q(t|\boldsymbol{\theta}) u_{q}(t) \right] \textrm{d}t.
\end{eqnarray}
Then the square $\mathbb{L}_{2}$ norm of the differential equation in (\ref{ODE}) with $x(t)$ replaced 
by the basis function expansion in (\ref{BE2}), can be written as 
\begin{eqnarray}\label{MatODE}
\nonumber
\textbf{L}(\boldsymbol{\theta} )&=& \sum_{k=1}^{K} \sum_{j=1}^{K} \int_{t_{1}}^{t_{N}}\left[\textbf{c}_{k} 
D^{p}\phi_{k}(t) +
\sum_{r=0}^{p-1} \beta_r(t|\boldsymbol{\theta})  \textbf{c}_{k} D^{r}\phi_{k}(t) - \sum_{q=1}^{Q}   
\alpha_q(t|\boldsymbol{\theta}) u_{q}(t) \right] \times \\
\nonumber
&& \quad \quad \, \left[\textbf{c}_{j} D^{p}\phi_{j}(t) +
\sum_{r=0}^{p-1} \beta_r(t|\boldsymbol{\theta})  \textbf{c}_{j} D^{r}\phi_{j}(t) - \sum_{q=1}^{Q}   
\alpha_q(t|\boldsymbol{\theta}) u_{q}(t) \right] \textrm{d}t,\\
&=&  
\sum_{k=1}^{K} \sum_{j=1}^{K} \left[ \textbf{c}_{k}\textbf{R}_{k,j}(\boldsymbol{\theta})\textbf{c}_{j} + 
2\textbf{c}_{k}\textbf{S}_{k}(\boldsymbol{\theta}) \right]+ 
\int_{t_{1}}^{t_{N}} \left[  \sum_{q=1}^{Q}  \alpha_q(t|\boldsymbol{\theta}) u_{q}(t) \right]^{2} \textrm{d}t.
\end{eqnarray}
The coefficients $\hat{\textbf{c}}(\boldsymbol{\theta},\rho)$ for fixed $\boldsymbol{\theta}$ and $\rho$ 
are obtained by minimising the penalised least squares criterion
\begin{eqnarray}\label{Mopt2}
\nonumber
J(\textbf{c}|\boldsymbol{\theta},\rho) & = & 
\frac{(1-\rho)}{N}  
\left[\textbf{y} - \boldsymbol{\Phi}  \textbf{c} \right]^{T}\left[\textbf{y} - \boldsymbol{\Phi}  \textbf{c} 
\right]+ \frac{\rho}{(t_{N}-t_{1})} \left[ 
\textbf{c}^{T}\textbf{R}(\boldsymbol{\theta})\textbf{c}+2\textbf{c}^{T}\textbf{S}(\boldsymbol{\theta})\right]
 \\
&&+ \frac{\rho}{(t_{N}-t_{1})}  \int_{t_{1}}^{t_{N}} \left[  \sum_{q=1}^{Q}  \alpha_q(t|\boldsymbol{\theta}) 
u_{q}(t) 
\right]^{2} \textrm{d}t,
\end{eqnarray}
\noindent where $\textbf{y}$ is a vector of length $N$ containing the measured observations, 
$\boldsymbol{\Phi}$ is an $N\times 
K$ matrix containing the elements $\phi_{k}(t_{i})$ for $i=1,\ldots,N$ and $k=1,\ldots,K$ and 
$\textbf{c}$ is a vector of length $K$ containing the coefficients of the basis functions. Equation 
(\ref{Mopt2}) combines two sources of information about $x(t),$ its fidelity to the data, as measured 
by the residual sum of squares in the first term in (\ref{Mopt2}), and its adherence to the linear 
differential equation in (\ref{ODE}), as quantified by the $\mathbb{L}_{2}$ norm of (\ref{ODE}) in 
the second and third term in (\ref{Mopt2}). To facilitate a comparable scale the fist term in (\ref{Mopt2}) 
is divided by $N$ to obtain the average of the squared residuals and the second and third term in 
(\ref{Mopt2}) is 
divided by $(t_{N}-t_{1})$ to obtain the average of the adherence to the linear differential equation. The 
regulating parameter $\rho$ can now be defined within the domain $[0,1)$. If $\rho=0$ then the 
corresponding estimated function, $\hat{\textbf{x}}=\boldsymbol{\Phi}  \hat{\textbf{c}},$ is the least squares approximation of the data and 
hence does not depend on the differential equation. However, as $\rho \rightarrow 1,$ minimising 
(\ref{Mopt2}) is equivalent to minimising (\ref{MatODE}). If (\ref{MatODE}) is approximately 
zero, $\hat{\textbf{x}}$ is an approximation of the solution of (\ref{ODE}).  The coefficient values that 
minimise (\ref{Mopt2}) with respect to $\textbf{c}$ for fixed $\boldsymbol{\theta}$ and $\rho$ are given 
analytically by
\begin{equation}\label{Copt}
\hat{\textbf{c}}(\boldsymbol{\theta},\rho) = 
\left[\frac{(1-\rho)}{N}\boldsymbol{\Phi}^{T}\boldsymbol{\Phi} + 
\frac{\rho}{t_{N}-t_{1}} \textbf{R}(\boldsymbol{\theta})\right]^{-1}
\left[\frac{(1-\rho)}{N} \boldsymbol{\Phi}^{T}\textbf{y} - 
\frac{\rho}{t_{N}-t_{1}} \textbf{S}(\boldsymbol{\theta})\right].
\end{equation}
See the supplementary material for the full derivation of (\ref{Copt}). 

The estimated parameters of the differential equation $\hat{\boldsymbol{\theta}}$ for fixed $\rho$ are 
obtained by minimising a dynamic model-fitting criterion
\begin{eqnarray}\label{thetaopt}
H(\boldsymbol{\theta}|\rho)&=& 
\left[\textbf{y}-\boldsymbol{\Phi}\hat{\textbf{c}}(\boldsymbol{\theta},\rho)
\right]^{T}\left[\textbf{y}-\boldsymbol{\Phi}\hat{\textbf{c}}(\boldsymbol{\theta},\rho)
\right],
\end{eqnarray}
where $\hat{\textbf{c}}(\boldsymbol{\theta},\rho)$ is given in (\ref{Copt}). Numerical optimisation 
methods such as Gauss-Newton can be used to minimise (\ref{thetaopt}) with 
respect to $\boldsymbol{\theta}$ for fixed $\rho$. The gradient of $H(\boldsymbol{\theta}|\rho)$ is 
required for the Gauss-Newton algorithm and is given analytically by
\begin{eqnarray} \label{cgradient}
\frac{\textrm{d} H(\boldsymbol{\theta}| \rho) }{\textrm{d} \boldsymbol{\theta}} &=& 
-2\left[\textbf{y}-\boldsymbol{\Phi}\hat{\textbf{c}}(\boldsymbol{\theta},\rho)\right]^{T}\boldsymbol{\Phi}\frac{\textrm{d}
	\hat{\textbf{c}}(\boldsymbol{\theta},\rho)}{ \textrm{d} \boldsymbol{\theta}}.
\end{eqnarray}
See the supplementary material for the formulae for evaluating  $\frac{\textrm{d} 
	\hat{\textbf{c}}(\boldsymbol{\theta},\rho)}{\textrm{d} \boldsymbol{\theta}}$. 

\subsection{The iterative scheme to acquire an optimal estimate of $\boldsymbol{\theta}$}
\label{subsec:rhoiteration}

The regulating parameter $\rho$ controls the complexity of the surface in (\ref{thetaopt}). For low 
values of $\rho$,  $\hat{\textbf{x}}$ is a least-squares 
approximation of the data and $H(\boldsymbol{\theta}|\rho)$ is convex. Thus, the minimum of 
$H(\boldsymbol{\theta}|\rho)$ with respect to $\boldsymbol{\theta},$ 
is easy for the Gauss-Newton algorithm to locate. Low values
of $\rho$ do not require $\textbf{L}(\hat{\boldsymbol{\theta}})$ to be small.  Consequently 
$\hat{\textbf{x}}$ is not a satisfactory approximation of the solution of the differential equation in 
(\ref{ODE}) and $\hat{\boldsymbol{\theta}}$ is an inaccurate estimate of $\boldsymbol{\theta}$. 
For 
high values of $\rho$, 
$H(\boldsymbol{\theta}|\rho)$ is a non-convex surface 
with flat plains, ripples and a long narrow ridge around the global minimum. In this instance, unless the 
initial estimate for $\boldsymbol{\theta}$ is within the narrow basin of attraction of the global minimum 
the Gauss-Newton algorithm will converge to a local minima.  High values of $\rho$ require 
$\textbf{L}(\hat{\boldsymbol{\theta}})$ to be small and therefore $\hat{\textbf{x}}$ is a better 
approximation to the solution of the differential equation in (\ref{ODE}) and $\hat{\boldsymbol{\theta}}$ 
is a more precise estimate of $\boldsymbol{\theta}$.
  
\begin{figure}[!h]
	\centering
	\includegraphics[width=0.8\textwidth]{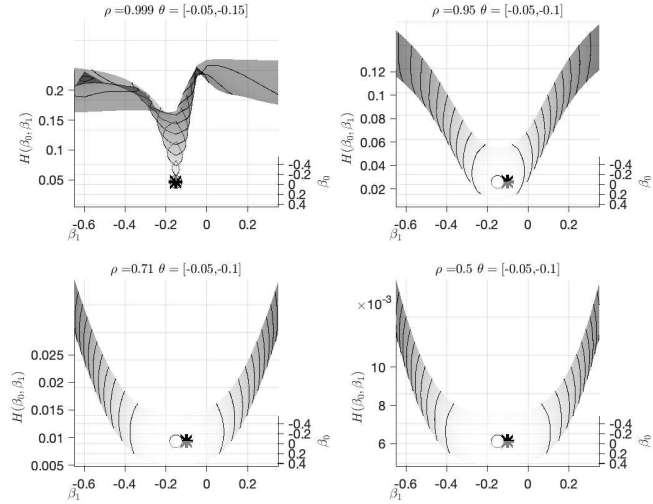}
	\caption{The optimisation surface $H(\boldsymbol{\theta}|\rho)$ for the simulation of the head 
		acceleration analysis discussed in 
		Section (\ref{sec:intro}) with $\rho=0.5, 0.71, 0.95$ and $\rho=0.99$ with $\beta_0$ 
		ranging from $-0.55,0.45$ and $\beta_1$ ranging from $-0.65,0.35$. In each figure the global 
		minimum is identified by the large dot and the local minima is identified by the asterisk.}\label{Opt}
\end{figure}

To illustrate this we examine the estimates of $\boldsymbol{\theta}$ obtained by minimising 
$H(\boldsymbol{\theta}|\rho)$ with respect 
$\boldsymbol{\theta}$ for various values of $\rho$ for a simulated data set. The 
data are obtained by evaluating the analytic solution of (\ref{HIE}) with 
$\boldsymbol{\theta}=[-0.05,-0.15,0.39]$ at $101$ equally spaced points within the 
domain $[0,60]$ and adding a vector of $101$ independently normally distributed random variables 
with mean $0$ and $\sigma = 0.05.$ Figure (\ref{Opt}) shows the surfaces and contours of 
$H(\boldsymbol{\theta}|\rho)$ for $\boldsymbol{\theta}=[\beta_0,\beta_1,0.39],$ with $\beta_0$ 
ranging from $-0.55$ to $0.45$ and $\beta_1$ ranging from $-0.65$ to $0.35$ with $\rho=0.99, 0.95, 
0.71$ and $0.50$. The circle is the true value for $(\beta_0,\beta_1) = (-0.05,-0.15)$ and the asterisk 
represents the minimum of $H(\boldsymbol{\theta}|\rho)$ for the respective $\rho$. For 
$\rho=0.50$ the 
surface is quadratic, but the minima $\hat{\boldsymbol{\theta}}=[-0.05,-0.10,0.39],$ is not an 
accurate estimate of $\boldsymbol{\theta}$. For $\rho=0.99$ the surface has flat plains, 
ripples and a long narrow ridge which is difficult for gradient decent methods to navigate, but the 
minima $\hat{\boldsymbol{\theta}}=[-0.05,-0.15,0.39],$ is an accurate estimate of 
$\boldsymbol{\theta}$. 
%%%%%%%%%%%%%%%%%%%%%%%%%%%%%%%%%%%%%%%%%%%%%%%%%%%%%%%%%%%%%%%%%%%%%%%%%%%%%%%%%%%%%%%%%%%%%%%%%
%%%%%%%%%%%%%%%%%%%%%%%%%%%%%%%%%%%%%%%%%%%%%%%%%%%%%%%%%%%%%%%%%%%%%%%%%%%%%%%%%%%%%%%%%%%%%%%%%
Figure (\ref{Opt}) also shows the changes in the topology of $H(\boldsymbol{\theta}|\rho)$ with respect 
$\rho$. As $\rho$ reduces the non-convex surface, which is a delta function 
located at the global minimum, illustrated by $H(\boldsymbol{\theta}|0.99)$, is smoothed to a convex 
surface with a flat area around the global minimum as illustrated by $H(\boldsymbol{\theta}|0.5)$. Thus, 
the regulating parameter $\rho$ controls the prominence of local and global minima and as such has the 
same role as the temperature parameter $T$ in simulated annealing described in Section (\ref{Sim_A}).

Data to linear dynamics, navigates the complicated topology of $H(\boldsymbol{\theta}|\rho)$ by 
starting with a relatively small value of $\rho$ (e.g. $\rho^{0}=0.04$) and an infeasible (exterior) point 
$\hat{\boldsymbol{\theta}}^{(0)}$ (e.g. $\hat{\boldsymbol{\theta}}^{(0)}=\left[.01,\ldots,.01\right]$) so 
that no steep valleys are present in the initial optimisation of $H( 
\boldsymbol{\theta} |\rho^{0})$. The difference between consecutive values of $\rho$ can be larger for 
$\rho<0.9$ for which $H( \boldsymbol{\theta} |\rho)$ is relatively convex. For $\rho>0.9$ the difference 
between consecutive values of $\rho$ must be small as the surface of $H( \boldsymbol{\theta} |\rho)$ 
can change substantially from one value of $\rho$ to the next. As a consequence, $H( 
\boldsymbol{\theta} |\rho)$ is minimised with logistic values of $\rho$ chosen so that the minimum of each 
$H( \boldsymbol{\theta} |\rho)$ is ``close" to the previous one. This will help to preclude difficulties in 
finding the global minimum of $H( \boldsymbol{\theta} |\rho)$ from one iteration to the next. Analogous 
to the many choices for the temperature ladder in simulated annealing, see 
\cite{stander1994temperature} for details, one could envisage many possible methods for reducing 
$\rho$ from one iteration to the next. Our approach proposed herein is a rather conservative approach 
and we acknowledge that an optimal reduction of $\rho$ is an area for future research.  The estimation 
procedure stops when the estimated parameters converge. The details of the Data to linear dynamics 
iterative scheme are below:  
\begin{description}
	\item [\textit{Step} 1]Specify the initial values: $\hat{ \boldsymbol{\theta}}^{0}$ (e.g. 
	$\hat{\boldsymbol{\theta}}^{(0)}=\left[0.01,\ldots,0.01\right]$) and  $\gamma^{0}= -4$.
	\item[\textit{Step} 2:]
	Let $\hat{\rho}^{u} = \frac{\exp(\gamma^{u})}{1 + \exp(\gamma^{u})}$ and obtain 
	$\hat{\boldsymbol{\theta}}^{u}$ by minimising $H( \boldsymbol{\theta} |\hat{\rho}^{u})$ in 
	(\ref{thetaopt}) with respect to $\boldsymbol{\theta}$ using Gauss-Newton methods. The 
	initial values for $\boldsymbol{\theta}$ are $\hat{\boldsymbol{\theta}}^{u-1}$ and the gradient of $H( 
	\boldsymbol{\theta} |\hat{\rho}^{u})$ is given in (\ref{cgradient}) with $\rho$ replaced by 
	$\hat{\rho}^{u}$.
	\item[\textit{Step} 3:] If the relative change between the local minimum of the objection function $H( \boldsymbol{\theta} |\rho)$ for two successive iterates, $\frac{H( \boldsymbol{\theta} 
	|\hat{\rho}^{u})-H( 
	\boldsymbol{\theta} |\hat{\rho}^{u-1})}{H( \boldsymbol{\theta} |\hat{\rho}^{u-1})},$ is smaller than $\epsilon_0$ (e.g. $\epsilon_0 = 0.2$) then let
	$\gamma^{u}=\gamma^{u-1}+1$ otherwise 
	$\gamma^{u}=\gamma^{u-1}+\frac{\gamma^{u}-\gamma^{u-1}}{2}$. 
	\item[	\textit{Step} 4: ] If the distance between the estimated parameters of the differential equation 
	for two successive iterates, $\hat{\boldsymbol{\theta}}^{u} -\hat{\boldsymbol{\theta}}^{u-1}$ is 
	smaller than $\epsilon_1$ 
	(convergence tolerance for selecting an optimal $\hat{\boldsymbol{\theta}}$ e.g. $\epsilon_1 = 
	10^{-4}$) stop. 
\end{description}
Let $u_{max}$ be the value of $u$ when the iterative scheme stopped. The iterative scheme produces 
$\hat{ \boldsymbol{\theta}}=\hat{ 
	\boldsymbol{\theta}}^{u_{max}}$ the estimated parameters of the differential equation that best 
	approximate the data. Substituting $\hat{\boldsymbol{\theta}}$ and $\hat{\rho}=\hat{\rho}^{u_{max}}$ 
	into (\ref{Copt}) produces an estimate of the coefficients of the basis function expansion 
	$\hat{\textbf{c}}(\hat{\boldsymbol{\theta}},\hat{\rho})$. The approximated solution of the differential 
	equation is $\hat{\textbf{x}}=  \boldsymbol{\Phi}\hat{\textbf{c}}(\hat{ \boldsymbol{\theta}},\hat{\rho})$ 
	and its degrees of freedom are
\begin{eqnarray}\label{degree_free}
\nonumber
\hat{\textrm{df}} &=& 
\textrm{trace}\left(2\boldsymbol{\Phi}\textbf{M}(\hat{\boldsymbol{\theta}},\hat{\rho})\boldsymbol{\Phi}^{T}\boldsymbol{\Phi}\textbf{M}(\hat{\boldsymbol{\theta}},\hat{\rho})\boldsymbol{\Phi}^{T}\boldsymbol{\Phi}\textbf{M}(\hat{\boldsymbol{\theta}},\hat{\rho})'\boldsymbol{\Phi}^{T}
\right), 
\end{eqnarray}
where the $K \times K$ matrix $\textbf{M}(\boldsymbol{\theta},\rho) = 
\left[\frac{(1-\rho)}{N}\boldsymbol{\Phi}^{T}\boldsymbol{\Phi} + 
\frac{\rho}{t_{N}-t_{1}} \textbf{R}(\boldsymbol{\theta})\right]^{-1}
$. Here degrees of freedom refers to the effective dimensionality of $\hat{\textbf{x}},$ as $\hat{\rho} 
\rightarrow 1$ it tends to the dimensionality of the solution space for the differential equation.

\subsection{Approximating the sampling variation for $\hat{\boldsymbol{\theta}}$ and 
$\hat{\textbf{x}} $}
\label{subsec:inference}

Assuming that $\textbf{y}$ is normally distributed with variance $\sigma^{2}_{y}$. The conditional 
sampling variance of the estimated parameters of the differential equation can be 
approximated using the delta method \citep{BatesWatts1988}:
\begin{eqnarray} \label{thetavariance}
\nonumber
\textrm{Var}[\hat{\boldsymbol{\theta}} | \rho ] &\approx& \hat{\sigma}^{2}_{y} \left( \frac{ 
	\textrm{d} 
	\hat{\boldsymbol{\theta}}  }{ \textrm{d} \textbf{y}} \right) \left(  \frac{ \textrm{d} 
	\hat{\boldsymbol{\theta}}  }{ 
	\textrm{d} \textbf{y}} \right)^{T},
\end{eqnarray}
where $\hat{\sigma}^{2}_{y} = 
\frac{\|\textbf{y}-\boldsymbol{\Phi}\hat{\textbf{c}}(\hat{\boldsymbol{\theta}},\hat{\rho})
	\|^{2}}{(N-\hat{\textrm{df}})}$. The formula for evaluating  $\frac{\textrm{d} \hat{\boldsymbol{\theta}} 
	}{ \textrm{d} \textbf{y}}$ is given in the supplementary material.

The point-wise conditional sampling variance of the estimated solution of the differential equation 
evaluated at the 
data points $\hat{\textbf{x}}$ is also approximated using the delta method:
\begin{eqnarray}\label{xoftvariance}
\nonumber
\textrm{Var}[\hat{\textbf{x}} | \rho] &\approx& 
\hat{\sigma}^{2}_{y} \boldsymbol{\Phi}^{T} 
\left( \frac{ \textrm{d} \hat{\textbf{c}}(\hat{\boldsymbol{\theta}},\hat{\rho})  }{ \textrm{d} 
	\textbf{y}}^{T} \right)      
\left( \frac{ \textrm{d} \hat{\textbf{c}}(\hat{\boldsymbol{\theta}},\hat{\rho})  }{ \textrm{d} 
	\textbf{y}}  \right)
\boldsymbol{\Phi}.
\end{eqnarray}
The formula for evaluating  $\frac{ \textrm{d}  
	\hat{\textbf{c}}(\hat{\boldsymbol{\theta}},\hat{\rho})  }{ \textrm{d} \textbf{y}}$ is given in the 
	supplementary material.
	
\section{A differential equation for modelling head acceleration}\label{subsec:TBI}

\begin{figure}[!h]
	\begin{center}
		\includegraphics[width=.8\textwidth]{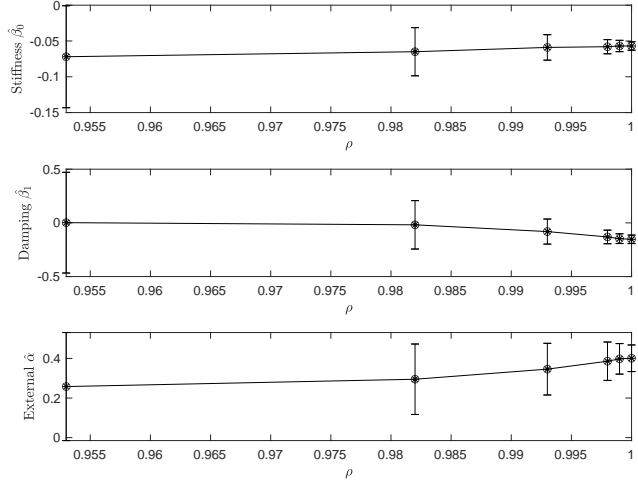}
		\caption{The values of the three parameters and their approximated 95\% confidence intervals for 
		the head impact data over values of $\rho$ 
			converging to 0.99.}
		\label{moto_flow}
	\end{center}
\end{figure}

\begin{figure}[!h]
	\centering
	\includegraphics[width=.8\textwidth]{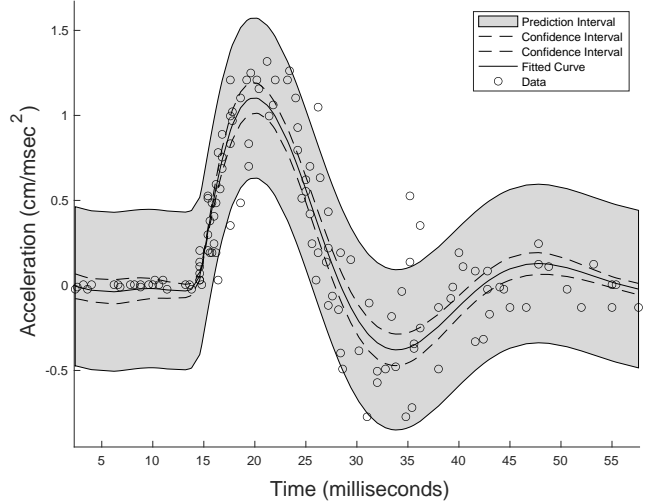}
	\caption{ The accelerometer readings of the brain tissue before and after a blow to the cranium are 
	indicated by the circles. The fitted curve produced by Data2LD with $\hat{\rho} = 0.99$ (solid line), 
	the approximated 
		95\% point-wise confidence interval for the curve (dashed line) and the approximated 95\% 
		point-wise prediction interval for the 
		curve (grey region).\label{Head_Imp}}
\end{figure}

We used three order one B-splines over the knots [0, 14, 15, 56] with coefficient vector [0,1,0] to 
represent the unit pulse function $u(t)$ in (\ref{HIE}). For the basis expansion of $x(t)$ we used order 
five B-spline functions, which by their nature have discontinuous third derivatives if all knots are 
singletons. The unit pulse function $u(t)$ is discontinuous, which implies a discontinuity in $D^2 x(t)$.
To achieve curvature discontinuity at the impact point and at that point plus one, we placed three knots 
at these locations. We put no knots between the first observation and the impact point, where the data 
indicate a flat trajectory and eleven equally spaced knots between the impact point plus one and the 
final observation time. We estimated the coefficients $\textbf{c}$ and the parameters 
$\boldsymbol{\theta} = \left[\beta_{0},\beta_{1},\alpha\right]$ in (\ref{HIE}) using Data2LD. Figure 
(\ref{moto_flow}) shows how the three parameters of the differential equation in (\ref{HIE}) and their 
approximated confidence intervals vary as $\rho$ increased to $0.99$.  
%%%%%%%%%%%%%%%%%%%%%%%%%%%%%%%%%%%%%%%%%%%%%%%%%%%%%%%%%%%%%%%%%%%%%%%%%%%%%%%%%%%%%%%%%%%%%%%%%%%
%%%%%%%%%%%%%%%%%%%%%%%%%%%%%%%%%%%%%%%%%%%%%%%%%%%%%%%%%%%%%%%%%%%%%%%%%%%%%%%%%%%%%%%%%%%%%%%%%%%
As shown in Figure (\ref{moto_flow}) when the influence of the differential equation increases to the point 
where it is the primary determinant of the parameters, the parameter values stabilise, and the 
approximated confidence intervals reduce. The final parameter estimates with 95\% confidence intervals 
are, $\hat{\beta}_{0} =  -0.057 \pm 0.005$ for the stiffness, $\hat{\beta}_{1} = -0.15  \pm 0.03$ for the 
damping and $\hat{\alpha} = 0.40 \pm 0.06$ for the force from the unit pulse function. Implying that the 
acceleration is an under-damped process; after the blow to the cranium, the acceleration will oscillate 
with a decreasing amplitude that will quickly decay to zero. The parameters of the differential equation 
suggest that it will take approximately 66 milliseconds after impact for the average acceleration to return to zero. 
The estimated function $\hat{x}$ has an effective degrees of freedom that is equal to $2.45$, and a 
root mean squared error that is $0.05.$ Figure (\ref{Head_Imp}) shows the accelerometer readings of 
the brain tissue, the fitted curve produced by Data2LD (solid line), the approximated 95\% point-wise 
confidence interval for the fitted curve (dashed line) and the approximated 95\% point-wise prediction 
interval for the fitted curve (grey band).
	
The differential equation captures the trend in the acceleration of the brain tissue. It conveys that the acceleration peaks at approximately 6.2 milliseconds after impact and troughs at around 19.8 milliseconds after impact. 	

\section{Simulation Study: head acceleration model}
\label{sec:simulation}

Consider the differential equation
\begin{equation}\label{Sim1}
D^{2}x(t)= -0.05 x(t) -0.15Dx(t)   +  0.39u(t),
\end{equation}
where $u$ is one for  $14 \leq t \leq 15$ and zero otherwise and the initial conditions are $x(0)=0$ and $Dx(0)=0$. Let $\textbf{x}$ be the analytic solution of (\ref{Sim1}), evaluated at $N$ equally-spaced observations over the domain $[0,60].$ The data $\textbf{\textsf{y}}$ are generated by  
$
\textbf{\textsf{y}}=\textbf{x}+\boldsymbol{\varepsilon}
$
where $\boldsymbol{\varepsilon}$ is a vector of $N$ independent normally distributed random values with mean $0$ and standard deviation $\sigma \times \textrm{range}(\textbf{x}).$  One thousand simulated samples 
are generated for $\sigma = 0.01, 0.05, 0.10$, and for sample sizes $N = 21,51,101$. The basis 
functions for approximating $x$ and $u$ are set up as described in Section \ref{subsec:TBI}. 

To implement SFT we let the priors for $\beta_{0},\beta_{1}$ and $\alpha$  be normal distributions with mean $0$ and variance $1$. The prior for $\sigma^{2}$ was chosen to be $\frac{1}{\sigma^2}$. Four parallel chains were used with four different temperatures $\{10,100,1000,10000\}$. We ran fifty thousand parallel MCMC chains and each chain was initialised with the same values. As suggested in \cite{kirkpatrick1983optimization} for SA we set the initial temperature to $T=100$ and reduced it using a Boltzmann schedule.  

Table (\ref{GPE1}) provides the root mean squared error of the estimated parameters $\hat{\boldsymbol{\theta}}$ with respect to the true parameters $\boldsymbol{\theta}_{\textrm{true}},$ $\textrm{RMSE}(\hat{\boldsymbol{\theta}})$ for the estimates obtained by Data2LD, simulated annealing (SA), smooth functional tempering (SFT), non-linear least squares (NLS) and parameter cascading (PC). The minimum $\textrm{RMSE}(\hat{\boldsymbol{\theta}})$ for each of the nine simulated data configurations involving three sample sizes and three levels of error is highlighted in grey.
\begin{table}[!h]
	\centering
	\caption{The root mean squared errors (RMSE) times 100, averaged over 1000 simulations for 
		$\hat{\beta}_{0}$, $\hat{\beta}_{1}$ and $\hat{\alpha}$ for the head acceleration analysis estimated by 
		Data2LD, simulated annealing (SA), non-linear least squares (NLS), smooth functional tempering (SFT) and parameter cascading (PC). The 
		minimum RMSE for each of the nine simulated data 
		configurations involving three sample sizes and three levels of error is highlighted in grey.} 
	\label{GPE1}
	\begin{tabular}{|l|lll|lll|lll|}
		\hline
		&\multicolumn{3}{|c|}{\textbf{$100 \times \textrm{RMSE}(\hat{\beta}_{0})$}} & \multicolumn{3}{|c|}{\textbf{$100 \times \textrm{RMSE}(\hat{\beta}_{1}) $}} &
		\multicolumn{3}{|c|}{ \textbf{$100 \times \textrm{RMSE}(\hat{\alpha})$}} \\
		\hline
		N & $21$ & $51$ & $101$ &  $21$ & $51$ & $101$ & $21$ & $51$ & $101$\\
		\hline
		& \multicolumn{3}{|c|}{$\sigma=0.10$ } &\multicolumn{3}{|c|}{$\sigma=0.10$ } &\multicolumn{3}{|c|}{$\sigma=0.10$ } \\
		\hline
		Data2LD  & \fcolorbox{lightgray}{lightgray}{0.25}  &  \fcolorbox{lightgray}{lightgray}{0.17}  &  \fcolorbox{lightgray}{lightgray}{0.11} & \fcolorbox{lightgray}{lightgray}{1.79}  &  \fcolorbox{lightgray}{lightgray}{1.19}  &  \fcolorbox{lightgray}{lightgray}{0.12} & \fcolorbox{lightgray}{lightgray}{4.12}  &  \fcolorbox{lightgray}{lightgray}{2.37}  &  \fcolorbox{lightgray}{lightgray}{1.69}  \\
		SA & 1.95   & 0.22   & 0.16 & 6.38   & 1.76   & 1.28 & 32.49   & 3.74   & 1.95 \\
		NLS & 0.61  &  0.30   & 0.23 & 4.63  &  2.05   & 1.86 & 13.19  &  4.05   & 4.71\\
		PC & 0.27   & 0.23   & 0.21 & 2.57 & 2.06   & 1.70 & 27.34   & 27.31   & 27.16\\   
		SFT & 0.26 & 0.18 &0.13 & 2.46 & 1.88 & 0.75 &  39.42 & 36.84 & 34.91 \\
		\hline 
		& \multicolumn{3}{|c|}{$\sigma=0.05$ } &\multicolumn{3}{|c|}{$\sigma=0.05$ } &\multicolumn{3}{|c|}{$\sigma=0.05$ } \\
		\hline
		Data2LD  & \fcolorbox{lightgray}{lightgray}{0.11} &\fcolorbox{lightgray}{lightgray}{0.07} &   \fcolorbox{lightgray}{lightgray}{0.06} &   \fcolorbox{lightgray}{lightgray}{0.86} & \fcolorbox{lightgray}{lightgray}{0.62} & \fcolorbox{lightgray}{lightgray}{0.40} & \fcolorbox{lightgray}{lightgray}{1.84} & \fcolorbox{lightgray}{lightgray}{1.02} &\fcolorbox{lightgray}{lightgray}{0.96}\\
		SA & 0.40   & 0.14   & 0.07  &  3.75   & 1.01  &  0.49   &25.44   & 1.89    &0.93\\
		NLS & 0.28  &  0.14  &  0.09 &   2.52 &   0.85  &  0.58  & 5.94   & 2.22    &1.85\\
		PC & 0.36   & 0.23  &  0.22  & 10.39 &    1.97   & 1.81  & 27.32   &27.11   &23.19\\
		SFT & 0.20   & 0.11   & 0.06  &  1.12   & 0.81 &   0.68 &  45.07 & 39.45   & 37.38\\
		\hline 
		& \multicolumn{3}{|c|}{$\sigma=0.01$ } &\multicolumn{3}{|c|}{$\sigma=0.01$ } &\multicolumn{3}{|c|}{$\sigma=0.01$ } \\
		\hline
		Data2LD  & \fcolorbox{lightgray}{lightgray}{0.03}  & \fcolorbox{lightgray}{lightgray}{0.02}    & \fcolorbox{lightgray}{lightgray}{0.01} & \fcolorbox{lightgray}{lightgray}{0.18} & \fcolorbox{lightgray}{lightgray}{0.10} & \fcolorbox{lightgray}{lightgray}{0.08} & \fcolorbox{lightgray}{lightgray}{0.03} & \fcolorbox{lightgray}{lightgray}{0.02} &
		\fcolorbox{lightgray}{lightgray}{0.01}\\
		SA & 17.34  &  0.02  &  0.01  &  6.41  &  0.18  &  0.15 & 4.53  & 0.38 & 0.31\\     
		NLS & 0.07  &  0.02  &  0.02  &  0.44  &  0.12  &  0.13  &  1.33  & 0.31  &0.56\\
		PC & 22.84  &  0.90  & 0.46  & 25.84  & 3.56  & 3.31  & 26.15 & 19.56  &17.02\\	
		SFT & 0.04  &  0.04  &  0.03   & 0.34    & 0.23    & 0.23  & 39.61   & 39.24   &38.48\\
		\hline				
	\end{tabular}
\end{table}
Data2LD had the best performance for all three parameters, and all three are well determined by the data with the RMSE's being less than .5\% of the parameter magnitudes. SA and SFT are expected to produce lower RMSEs relative to NLS and PC, respectively, as these approaches are designed to deal with the complex topology of the parameter space. While SFT yielded lower RMSEs relative to PC for 
$\beta_0$ and $\beta_1,$ it increased the RMSE for $\alpha$ across all configurations.  SFT and PC showed considerably higher $\textrm{RMSE}(\hat{\alpha})$ across all sample sizes relative to the other approaches.  This indicates that neither SFT nor PC adequately estimated the sharp change in the process due to the impact of $u(t)$. SA showed a substantial increase in RMSE relative to NLS for $N=21$. This suggests that SA does not perform well when the sample size of the data set is small. 

Table (\ref{GPE2}) provides the 95\% coverage probability of the estimated confidence intervals $\textrm{CP}(\hat{\boldsymbol{\theta}})$ for the estimates obtained by Data2LD, SA, SFT, NLS and PC. The coverage probability measures the proportion of the estimated confidence intervals for $\hat{\boldsymbol{\theta}}$ that contained the true parameters $\boldsymbol{\theta}_{\textrm{true}}$ over 1000 simulations. The closest to 95\% for each of the nine simulated data configurations involving three sample sizes and three levels of error is highlighted in grey.
%%%%%%%%%%%%%%%%%%%%%
\begin{table}[!h]
	\centering
	\caption{The 95\% coverage probability of the estimated confidence intervals averaged over 1000 simulations for 
		$\hat{\beta}_{0}$, $\hat{\beta}_{1}$ and $\hat{\alpha}$ for the head acceleration analysis estimated by 
		Data2LD, simulated annealing (SA), non-linear least squares (NLS), smooth functional tempering (SFT) and parameter cascading (PC).The coverage probability closest to 95\% for each of the nine simulated data configurations involving three sample sizes and three levels of error is highlighted in grey.} 	\label{GPE2}
	\begin{tabular}{|l|lll|lll|lll|}
		\hline
		&\multicolumn{3}{|c|}{\textbf{95\% $\textrm{CP}(\hat{\beta}_{0})$}} & \multicolumn{3}{|c|}{\textbf{95\% $\textrm{CP}(\hat{\beta}_{1})$}} &
		\multicolumn{3}{|c|}{\textbf{95\% $\textrm{CP}(\hat{\alpha})$}} \\
		\hline
		N & $21$ & $51$ & $101$ &  $21$ & $51$ & $101$ & $21$ & $51$ & $101$\\
		\hline
		& \multicolumn{3}{|c|}{$\sigma=0.10$ } &\multicolumn{3}{|c|}{$\sigma=0.10$ } &\multicolumn{3}{|c|}{$\sigma=0.10$ } \\
		\hline
		Data2LD  & \fcolorbox{lightgray}{lightgray}{88} &  \fcolorbox{lightgray}{lightgray}{90}  &  \fcolorbox{lightgray}{lightgray}{92} &   \fcolorbox{lightgray}{lightgray}{86}  & \fcolorbox{lightgray}{lightgray}{89} &   \fcolorbox{lightgray}{lightgray}{91}  &  \fcolorbox{lightgray}{lightgray}{87}  &  
		\fcolorbox{lightgray}{lightgray}{86}    & 
		\fcolorbox{lightgray}{lightgray}{87}\\
		SA &79    &99   &100    &39   & 78  &  85 &   25    &68  &  80\\
		NLS & 82  &  86  &  71   & 78  &  78    & 64    &53 &   61&    55\\
		PC & 55 & 61 & 61 & 41 & 52 & 57 & 0 & 0 & 0 \\
		SFT & 100 & 100 & 100 & 100 & 100 & 100 &100 &100 &100\\
		\hline
		& \multicolumn{3}{|c|}{$\sigma=0.05$ } &\multicolumn{3}{|c|}{$\sigma=0.05$ } &\multicolumn{3}{|c|}{$\sigma=0.05$ } \\
		\hline
		Data2LD  &	\fcolorbox{lightgray}{lightgray}{82}  &  91  &  \fcolorbox{lightgray}{lightgray}{92}   & \fcolorbox{lightgray}{lightgray}{81} &   \fcolorbox{lightgray}{lightgray}{91}  &  \fcolorbox{lightgray}{lightgray}{85} &   \fcolorbox{lightgray}{lightgray}{86}   & \fcolorbox{lightgray}{lightgray}{89}   & 
		\fcolorbox{lightgray}{lightgray}{86}\\
		SA & 76  &  \fcolorbox{lightgray}{lightgray}{93}  &  99 &   38  &  68  &  77 &   15  &  55  &  64\\
		NLS & 78  &  74 &   74  &  69 &   72  &  68 &   51  &  57 &   56\\
		PC & 76 & 51 & 38 & 30 &24 &17 &0 &0&3\\
		SFT & 100 & 100 & 100 & 100 & 100 & 100 &100 &100 &100\\
		\hline
		& \multicolumn{3}{|c|}{$\sigma=0.01$ } &\multicolumn{3}{|c|}{$\sigma=0.01$ } &\multicolumn{3}{|c|}{$\sigma=0.01$ } \\
		\hline
		Data2LD  &	\fcolorbox{lightgray}{lightgray}{82}  &  82 &   \fcolorbox{lightgray}{lightgray}{91}    &\fcolorbox{lightgray}{lightgray}{85} &   \fcolorbox{lightgray}{lightgray}{90} &   \fcolorbox{lightgray}{lightgray}{90} &   \fcolorbox{lightgray}{lightgray}{89}&    
		\fcolorbox{lightgray}{lightgray}{84}    & 
		\fcolorbox{lightgray}{lightgray}{94}\\
		SA & 66   & \fcolorbox{lightgray}{lightgray}{94}  &  98  &  19    &44 &   72 &    8  &  34    &43\\
		NLS &79  &  75 &   59 &   78    &80   & 79  &  54 &   61    &53\\
		PC & 0 & 21 & 91 & 0 & 13 &76 &0 &8&9\\
		SFT & 100 & 100 & 100 & 100 & 100 & 100 &100 &100 &100\\
		\hline
	\end{tabular}
\end{table}
In most cases, Data2LD obtained the most accurate approximation of the uncertainty associated with the estimate $\hat{\boldsymbol{\theta}}.$ PC substantially underestimated the coverage probability in all cases except for $\sigma=0.01$ and $N=101$. SFT overestimated the coverage probability in all cases.

Table (\ref{GPE3}) assess the accuracy of the solution of the differential equation by reporting $\textrm{RMSE}(\hat{\textbf{x}})$, the root mean squared error of the estimated solution of the differential equation $\hat{\textbf{x}}$ with respect to the true solution.  The minimum RMSE for each of the nine configurations is highlighted in grey. 
\begin{table}[!h]
	\centering
	\caption{The root mean squared errors (RMSE) times 100 averaged over 1000 simulations for the 
		estimated solution of the differential equation for the head acceleration analysis estimated by 
		Data2LD, simulated annealing (SA), non-linear least squares (NLS), smooth functional tempering (SFT) and parameter cascading (PC).  The minimum RMSE for each of the 
		nine simulated data configurations involving three sample sizes and three levels of error is highlighted 
		in grey.}\label{GPE3}
	\begin{tabular}{|l|lll|lll|lll|}
		\hline
		&\multicolumn{3}{|c|}{\textbf{$100 \times \textrm{RMSE}(\hat{\textbf{x}})$}} & \multicolumn{3}{|c|}{\textbf{$100 \times \textrm{RMSE}(\hat{\textbf{x}})$}} &
		\multicolumn{3}{|c|}{ \textbf{$100 \times \textrm{RMSE}(\hat{\textbf{x}})$}} \\
		\hline
		& \multicolumn{3}{|c|}{$\sigma=0.10$ } &\multicolumn{3}{|c|}{$\sigma=0.05$ } &\multicolumn{3}{|c|}{$\sigma=0.01$ } \\
		\hline
		N & $21$ & $51$ & $101$ &  $21$ & $51$ & $101$ & $21$ & $51$ & $101$\\
		\hline
		Data2LD  &  \fcolorbox{lightgray}{lightgray}{3.36} &   \fcolorbox{lightgray}{lightgray}{2.30}   & \fcolorbox{lightgray}{lightgray}{1.60}  &  \fcolorbox{lightgray}{lightgray}{1.82}  &  \fcolorbox{lightgray}{lightgray}{1.14}    &\fcolorbox{lightgray}{lightgray}{0.85}    &\fcolorbox{lightgray}{lightgray}{0.32}   & \fcolorbox{lightgray}{lightgray}{0.26}    &\fcolorbox{lightgray}{lightgray}{0.20}\\
		SA & 9.41   & 2.91   & 1.96  &  4.83   & 1.67    &1.06   & 5.43  &  0.33  &  0.24\\
		NLS& 5.28 &   2.88  &  2.27 &   3.19  &  1.36    & 1.18   & 0.75  &  0.29   & 0.29\\
		PC & 4.41   & 3.11  &  2.76  &  3.85  &  2.55  &  2.39  &  1.17  &  0.76  &  0.34\\
		SFT &7.67  & 5.33 &   2.94  &    3.81   & 2.22    & 1.56 &   0.65 &    0.45 &    0.30\\
		\hline 
	\end{tabular}
\end{table}
Data2LD has the best performance with the lowest 
RMSE for each of the nine configurations. SA showed an increase in RMSE relative to NLS for $N=21$ and $N=51.$ Indicating that SA provides an improvement in the estimate of the solution of the differential equation only when $N$ is large. SFT showed an increase in RMSE relative to PC for $\sigma=0.1$. Indicating that SFT  does not provide an improvement in the estimate of the solution of the differential equation unless $\sigma$ is small. \\

Data2LD, PC, NLS, SA and SFT had average computation times of 3.64, 36.65, 284.67, 1844.18 and 3677.57 seconds per simulation, respectively, executed in Matlab (2019a) on a 4 GHz iMac computer. Data2LD is over thousand times faster than SFT and over five hundred faster than SA.

\section{Discussion and Conclusions}
\label{sec:conclusion}

Dynamical systems can provide a conceptual understanding of how processes evolve, which can help guide their management and prediction or can simply provide a tractable, flexible and parsimonious model of the processes. The parameters of a dynamical system determine the interrelationships between the processes which describe how these objects be it physical, engineering or demographic behave. These parameters are often unknown and must be estimated from the observed data. The most popular approaches for parameter estimation for dynamical systems are smooth functional tempering (SFT), parameter cascading (PC), simulated annealing (SA) and non-linear least squares (NLS).

The NLS and PC approach involves obtaining the minimum of 
(\ref{NLS}) and (\ref{theta}) with respect to the parameters' of the differential equation. These parameter spaces can exhibit complex topology including multi-modality, ripples and narrow ridges and as such, can be difficult to navigate. SA and SFT are popular approaches for finding the global minimum of (\ref{NLS}) and (\ref{theta}). As shown in \cite{gonzalez2006parameter, Campbell2012} and herein, SA and SFT often provide improved estimates of the parameters and the solution of the differential equation relative to NLS and PC. However, both SA and SFT are very computationally expensive and do not provide an adequate estimate of the uncertainty associated with the estimated parameters of the differential equation.  

We propose Data2LD a version of the PC approach that has been tailored for linear systems. First, we reduce the complexity of the PC estimation procedure, which has the advantages of speed and ease of use. Then analogous to SA, we propose an iterative scheme to overcome the topological difficulties in minimising the data misfit. One of the primary benefits of this algorithm is that it facilitates accurate and stable estimation of the solution, $x(t),$ and the parameters, $\boldsymbol{\theta}$ defining $\beta_{r}(t|\boldsymbol{\theta})$ and $\alpha_{q}(t|\boldsymbol{\theta})$ in (\ref{ODE}). 

We compared Data2LD with the popular existing approaches, namely SFT, PC, SA and NLS. In terms of statistical measures of performance such as root-mean-squared error for parameters estimates and the estimates of the solution of the differential equation, our simulations suggest an advantage for Data2LD. For large sample sizes, Data2LD and SA have a similar estimation accuracy with Data2LD having a computational advantage of about five orders of magnitude. SA does not perform well when the sample size of the data set is small. PC and SFT has difficulty estimating the sharp change in the solution at the impact point resulting in poor estimates of the parameters' of the differential equation.

We are extending Data2LD to linear dynamic systems along with data observed over space and time where processes can be denoted by a set of linear partial differential equations such as reaction-diffusion-transport family.

A Matlab package with source code and datasets for the examples presented in this article is available at \url{https://github.com/mcareyucd/Data2LD-Matlab}.  An R-package ``Data2LD" that contains functions for using differential equations as modelling objects can be obtained from CRAN at (\url{https://cran.r-project.org/web/packages/Data2LD/index.html}).

	%We  would like to thank the reviewers for their careful reading of our manuscript and their insightful comments and suggestions.

\section*{References}

\bibliography{mybibfile}

\end{document}